\documentclass[useAMS,usenatbib]{mn2e}

\usepackage{graphicx}
\usepackage{epsfig}
\usepackage{deluxetable}
\usepackage{longtable}
\usepackage{lscape}
\usepackage{txfonts}

\title[Nuclear clusters in late-type dwarf galaxies]{Globular cluster 
systems in nearby dwarf galaxies - II. Nuclear star clusters and 
their relation to massive Galactic globular clusters.\thanks{This 
study is based on archival data of the NASA/ESA {\it Hubble Space 
Telescope}, which is operated by AURA, Inc., under NASA contract 
NAS 5--26555.}}

\author[I.\,Y.\,Georgiev et al.]{{\Large Iskren Y. Georgiev$^1$\thanks{E-mail: 
iskren@astro.uni-bonn.de}, Michael Hilker$^2$, Thomas H. Puzia$^3$, 
Paul Goudfrooij$^4$ and Holger Baumgardt$^1$}\\
$^1$Argelander Institut f\"{u}r Astronomie der Universit\"{a}t Bonn, Auf dem H\"{u}gel 71, D-53121 Bonn, Germany\\
$^2$European Southern Observatory, 85748 Garching bei M\"unchen, Germany\\
$^3$Plaskett Fellow, Herzberg Institute of Astrophysics, 5071 West Saanich Road, Victoria, BC V9E 2E7, Canada\\
$^4$Space Telescope Science Institute, 3700 San Martin Drive, Baltimore, MD 21218, USA
}

\begin{document}

\date{Accepted 2009 March 12}

\pagerange{\pageref{firstpage}--\pageref{lastpage}} \pubyear{2009}

\maketitle

\label{firstpage}

\begin{abstract}
We compare nuclear globular clusters (nGCs) in dwarf galaxies and 
Galactic GCs with extended (hot) horizontal branches (EHBs-GCs) in order 
to test the suggested external origin of the latter and the conditions 
at which GC self-enrichment can operate. Using luminosities and structural 
parameters of nGCs in low-mass (mainly late-type) dwarf galaxies from 
HST/ACS imaging we derive the present-day escape velocities ($\upsilon_{\rm esc}$) 
of stellar ejecta to reach the cluster tidal radius and compare them with those of EHB-GCs. We show that 
nGCs in dwarf galaxies are very similar in their photometric and structural 
properties (colour $\langle V-I\rangle=0.9$, magnitudes $\langle M_V\rangle<-9$\,mag, 
ellipticities $\langle\epsilon\rangle=0.11$) to EHB-GCs. The nGCs populate the 
same $M_V$ vs. $r_{\rm h}$ region as EHB-GCs, although they do not reach the 
sizes of the largest EHB-GCs like $\omega$\,Cen and NGC\,2419. We argue that 
during accretion the $r_{\rm h}$ of an nGC could increase due to significant 
mass loss in the cluster vicinity and the resulting drop in the external 
potential in the core once the dwarf galaxy dissolves. For EHB-GCs, we find 
a correlation between the present-day $\upsilon_{\rm esc}$ and their metallicity 
as well as $(V-I)_0$ colour. The similar $\upsilon_{\rm esc}$ , $(V-I)_0$ 
distribution of nGCs and EHB-GCs implies that nGCs could also have complex 
stellar populations. The $\upsilon_{\rm esc}$--[Fe/H] relation could reflect 
the known relation of increasing stellar wind velocity with metallicity, 
which in turn could explain why more metal-poor clusters typically show 
more peculiarities in their stellar population than more metal-rich clusters 
of the same mass do. Thus the cluster $\upsilon_{\rm esc}$ can be used as 
parameter to describe the degree of self-enrichment. All our findings support 
the scenario in which Galactic EHB-GCs have originated in the centres of 
pre-Galactic building blocks or dwarf galaxies that were later accreted by 
the Milky Way.
\end{abstract}


\section{Introduction}

Several studies have
suggested that nuclear clusters of dwarf galaxies may be the progenitors, 
upon accretion, of the most massive globular clusters (GCs) in a galaxy 
\cite[e.g.,][]{Zinnecker88, Freeman93}. The most prominent case in our 
Galaxy is that of the GC M\,54, which sits in the core of the Sagittarius 
dwarf spheroidal (Sgr dSph) \cite[e.g.][]{Ibata97,Monaco05,Bellazzini08}.
The goal of the current study is 
to test this suggestion by means of a comparison of the properties of 
nuclear clusters in low-mass galaxies with those of the massive Galactic 
GCs. Recent studies based on deep observations of Galactic GCs challenged 
the traditional view of GCs being composed of a simple stellar population 
with homogeneous age and metallicity. Most massive Galactic GCs for which 
deep Hubble Space Telescope (HST) imaging is available exhibit multiple 
branches in their colour-magnitude diagrams such as $\omega$\,Cen 
\citep{Lee99,Hilker&Richtler00,Bedin04,Villanova07}, NGC\,2808 
\citep{Piotto07}, NGC\,1851 \citep{Milone08} and NGC\,6388 
\citep{Piotto08}. This implies that they host sub-populations 
with different chemical abundances and/or ages \cite[e.g. $\omega$\,Cen, ][]{Hilker04}. 
The presence of multiple stellar populations has recently stimulated 
studies on formation mechanisms to explain the complex stellar populations 
in GCs such as capturing of field stars \citep{Fellhauer06,Pflamm07}, 
repeated gas accretion \citep{Walcher05,Pflamm&Kroupa09}, merging of 
star clusters formed within the same GMC \citep{Mackey&Broby07}. 
The currently most favored mechanism to explain the multiple stellar 
populations and hot HB stars is the cluster self-enrichment from stars 
with enhanced mass-loss enriching the interstellar medium with He and other 
light elements, i.e. ($5-8M_{\odot}$) AGB stars \cite[e.g.][]{Ventura&DAntona08} 
or massive fast-rotating stars \citep{Prantzos06,Decressin07}. Helium 
enrichment from such sources is required to explain the sub-populations 
and the hot, core-helium and shell-hydrogen burning stars on 
the extension of the horizontal branch (HB) in GCs \citep{DAntona02,Piotto05,Maeder06}. 
Note however, that the \emph{late He flasher} scenario also can explain 
the extremely hot HB stars (blue hook stars) as stars which have experienced 
extreme mass-loss and late He-flash while descending to the white dwarf 
sequence \cite[e.g.][]{Brown01,Moehler07}.

Interestingly, NGC\,2419, the largest and among the most luminous GCs, 
has a very well defined population of hot stars on the extension of the 
HB, identical to $\omega$\,Cen \citep{Sandquist&Hess08, Dalessandro08}, 
but shows no clear evidence for multiple stellar populations and/or 
metallicity spread as deduced from its well defined red clump and narrow red 
giant branch in its HST/WFPC2 CMD \citep{Ripepi07}.

Typically the most complex clusters are among the most massive ones. 
Based on multivariate principle component analysis 
of various properties of Galactic GCs, \cite{Recio-Blanco06} showed that
there is a strong correlation between the HB extension (i.e., the maximum temperature reached on the HB) and the cluster 
total mass, accounting for up to $60\%$ of the HB variation. In addition, \cite{Dieball09} 
showed that ``blue hook'' stars (hotter than the EHB, $T_{\rm eff}\gtrsim31000$\,K, 
but with a similar FUV--NUV colour) are observed only in the most massive 
Galactic GCs. Therefore, the cluster mass is as important parameter for driving the HB morphology 
as the cluster metallicity and age. In this respect a more massive 
cluster (forming in a more massive giant molecular cloud, GMC) would be able to more efficiently 
retain stellar ejecta than a less massive one, thus leading to a 
higher degree of self-enrichment. Recently, \cite{Lee07} have shown 
that Galactic clusters with extended HBs (EHB-GCs), besides being 
among the most massive GCs in the Galaxy, form a kinematically distinct 
population {\it along} with the metal-poor ``Young Halo'' (YH) GCs. 
The fact that both the total orbital energy and the maximum distance 
from the Galactic plane correlate with the cluster metallicity for 
bulge/disc (BD) and non-EHB ``Old Halo'' (OH) GCs is consistent with 
the dissipational formation \citep{Eggen62} of those clusters together 
with the Galactic bulge and halo. The lack of such a correlation for YH-GCs and most of the 
EHB-GCs is consistent \citep{Lee07} with their formation in the 
cores of Galactic building blocks \citep{Searl&Zinn78} or in the 
nuclear regions of now defunct dwarf galaxies. 
If the most massive 
Galactic GCs had their origin as such nuclear clusters, they would 
have the ideal conditions for self-enrichment processes to take place. 
The retained stellar ejecta in the deep potential well can lead to the 
formation of stellar populations with different metallicities and/or ages.

One place to look for the progenitors of the peculiar Galactic EHB-GCs thus are the nuclear regions of galaxies 
harboring nuclear star clusters. Deep space-- and ground-based observations 
have recently established that many galaxies, irrespectively of their 
morphological type, contain star clusters at their nuclear regions. 
In particular, the frequency of nuclear clusters (NCs) in early-type 
spiral galaxies (Sa-Sc) with little or no bulge has been observed to 
be $\sim50\%$ \cite[e.g.][]{Carollo97}, in face-on late-type spirals 
(Scd-Sm) $\sim75\%$ \citep{Boker02} and in bulgeless edge-on spirals 
$\sim65\%$ \citep{Seth06}. These fractions should be regarded as 
lower limits as the detection of NCs depends on the spatial resolution 
of the observations. Spectra of NCs in late-type 
disc galaxies shows that they span a wide range in their luminosity 
weighted age from 10\,Myr to 
11\,Gyr with spectra best described by age-composite stellar populations 
and masses in the range $8-60\times10^6$\,M$_\odot$ \citep{Walcher05,Walcher06,Seth06} 
with the tendency of lower-mass NCs to be mainly found in later 
rather than earlier type spirals \citep{Rossa06}. The presence of NCs in 
dwarf elliptical galaxies has been established for a long time 
\cite[e.g.][]{Reaves83}. However, due to their high surface brightness 
and low photometric completeness of previous surveys the question of NC frequencies in 
ellipticals has been largely debated. 
Using deep HST/ACS 
imaging of 100 early-type galaxies in the Virgo cluster, \cite{Cote06} 
showed that at least $66\%$ (and as high as $87\%$) of the observed galaxies 
possess NCs.
So far there are no dedicated studies of NCs in late-type dwarf galaxies. 

The few existing observations of star clusters in late-type dwarfs 
show that the majority of the star clusters located in the 
nuclear regions (the inner 500\,pc) of nearby dIrrs (e.g., M\,82, 
NGC\,1705) are found to be predominantly young \cite[a few Myr,][]
{Ho&Filippenko96,Smith06,Westmoquette07}. 
However, old nuclear clusters have been reported in two dwarf spheroidal 
galaxies \citep{Puzia&Sharina08} as well as in seven dwarf irregular galaxies 
\citep{Georgiev08,Georgiev09}. The ages of those clusters are $>4$\,Gyr as 
inferred from their integrated spectra and broadband colours. 
In the following, we will refer to the old nuclear clusters in our sample 
as nuclear GCs (nGCs). 

In order to address the suggestion that the most massive and peculiar 
Galactic GCs could have had their origin in an environment similar to 
the nuclear regions of dIrr galaxies, we study their luminosities, colours, structural parameters and escape 
velocity to reach the cluster tidal radius. In Section\,\ref{Sect:Data} 
we briefly describe our dataset and the selection of nGCs, while in 
Section\,\ref{Sect:analysis} we perform the comparison between nGCs 
and EHB-GCs including their 
photometric and structural properties and their escape velocities. 
The main conclusions of this work are presented in Sect.\,\ref{Sect:Conclusions}.


\section{Data}\label{Sect:Data}

\subsection{Description}
This study is based on a two-band (F606W and F814W) archival HST/ACS 
dataset of nearby ($<12$\,Mpc), low-luminosity ($M_V>-16$\,mag, 
$L<5\times10^{8}L_{\odot}$), and mainly late-type dwarf galaxies which 
are located in low-density environments (in lose groups or in the halo regions 
of nearby galaxy groups). The dataset contains 30 dIrrs, 2 dEs, 
2 dSphs and 4 Sm dwarfs in which old GCs were detected.
Data reduction, photometry, completeness analysis, contamination 
and structural parameter measurements of all clusters are described 
in detail in \cite{Georgiev08,Georgiev09}, where we have also 
discussed in detail their colours, luminosities and structural 
parameters and how they compare with the Galactic GC sub-populations. 
Here we extend our study of the old GCs in those dwarf galaxies 
with particular attention to the brightest GCs located in their 
nuclear regions.

\subsection{Nuclear cluster selection}

\begin{figure*}
\epsfig{file=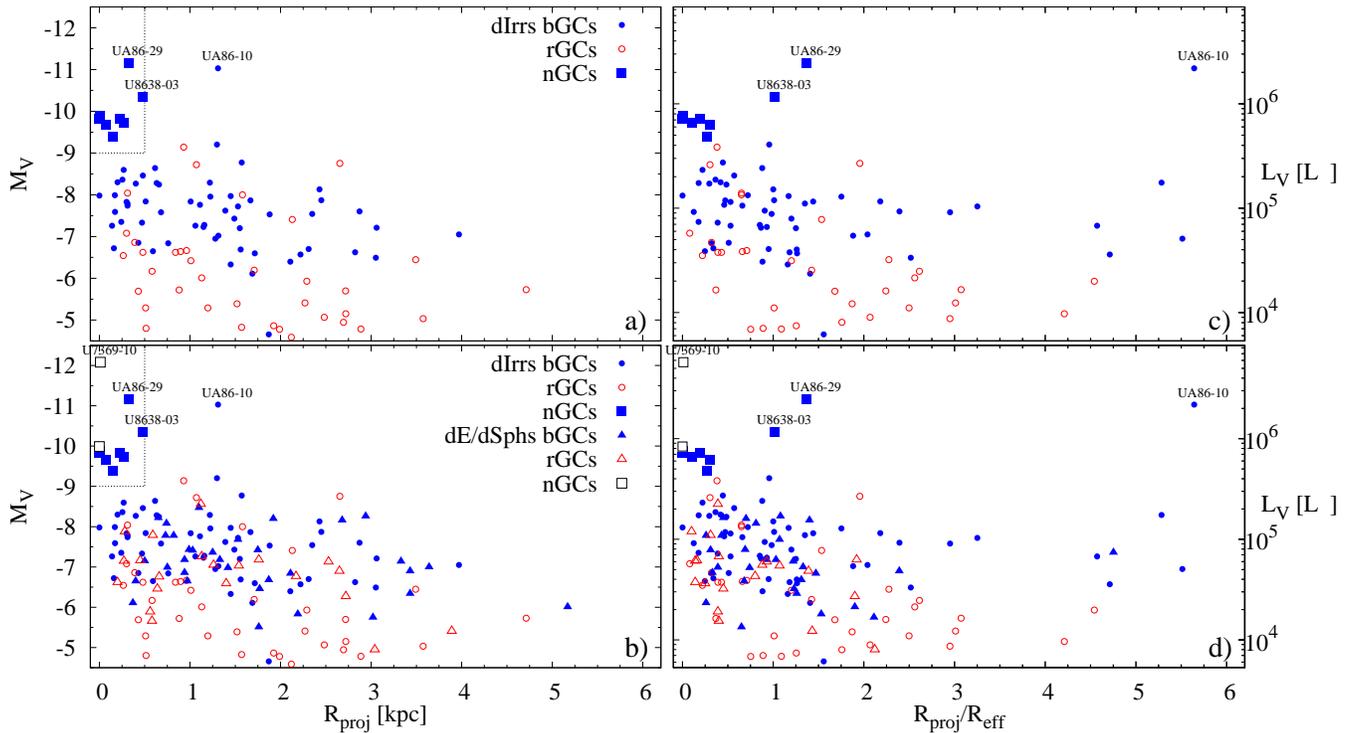, width=1\textwidth, bb=50 50 744 440}
\caption{Radial distribution as a function of the absolute 
magnitude/luminosity for GC candidates in low-mass dwarf galaxies. 
In panels a) and b) their projected galactocentric distances ($R_{\rm proj}$) 
are shown for dIrrs and dEs/dSphs, respectively. 
In panels c) and d) $R_{\rm proj}$ is normalized to the galactic effective 
radius $R_{\rm eff}=\sqrt{a\times b}$, i.e. the geometric mean of 
the galaxy semi-major and semi-minor axes. The dashed lines in 
panels a) and b) mark the region adopted for the selection of nuclear 
GCs (nGCs). With solid and open squares nGCs in dIrrs and dEs are 
indicated while solid and open circles and triangles mark clusters with colours 
bluer or redder than $V-I=1.0$\,mag in dIrrs and dEs, respectively.
\label{Fig.selection}}
\end{figure*}
The selection of nuclear clusters in low-mass galaxies is not a trivial 
task due to  
uncertainties which arise from the determination of 
the galaxy centre, especially for irregular galaxies where starburst 
regions are present. As presented in \cite{Georgiev09} we adopted the 
geometric centre of the isointensity contour at the $10\sigma$ level 
above the background as the galaxy centre. Recently, \cite{Swaters09} 
performed an HI kinematic analysis for a large sample of nearby 
late-type dwarf galaxies. We have two galaxies in common with their 
study: UGC\,1281 and NGC\,4163. The latter has a bright GC in 
the inner region. Unfortunately, \cite{Swaters09} were unable to derive 
the kinematical centre for this galaxy (and several other dwarfs in their 
sample) and adopted their photometric centre.

To define nuclear clusters we have constrained our selection 
criteria to clusters with $M_V < -9$\,mag and within a projected distance 
of $R_{\rm proj} < 500$\,pc from the galaxy centre (cf. Fig.\,\ref{Fig.selection}). 
As we will discuss later, there seems to be a clear division at this 
particular magnitude between nGCs and the rest of the GCs in those 
dwarfs. The value of the adopted projected distance was based on 
dynamical friction-related arguments. 
The dynamical friction time scale for a gravitating object 
on a Keplerian orbit in an isothermal halo is
$t_{\rm df}\propto(\sigma\times r_i^2)/M_{\rm Cl}$, where 
$\sigma$, $r_i$ and $M_{\rm Cl}$ are the local velocity dispersion (km/s), 
the initial distance from the centre (kpc) and the cluster mass ($M_\odot$), 
respectively \cite[e.g.][]{Binney&Tremaine08}. 
The adopted $R_{\rm proj}$ guarantees that a cluster with $10^4 M_\odot$ 
in a field of local $\sigma\sim15$\,km/s would spiral toward the centre 
of the host potential on a 
time scale of $t_{\rm df}\sim5$\,Gyr. Such a cluster we classify as 
nGC. For the case of M\,54 it has been shown by \cite{Monaco05} that 
this cluster might have spiraled from $r_i\lesssim4$\,kpc into the 
centre of the Sgr dSph within $t_{\rm df}\sim10$\,Gyr.
 
In Figure\,\ref{Fig.selection} we show the radial distribution of the 
GCs in our sample dwarf galaxies as a function of their luminosities. 
With different symbols are shown blue and red GCs in dIrrs or dEs/dSphs 
as indicated in the figure legend. Red GCs are those with $V-I>1.0$\,mag, 
which is the typical colour to separate metal-poor from metal-rich GCs 
in a galaxy. The left and right panels show distributions of the cluster 
luminosities as a function of the projected galactocentric distance 
(R$_{\rm proj}$) and R$_{\rm proj}$ normalized to the galaxy effective radius 
$R_{\rm eff}=\sqrt{a_{\rm eff}\times b_{\rm eff}}$, respectively. With dashed 
lines we have indicated the adopted selection region. This resulted in the 
selection of eight nGCs in seven dIrrs and two nGCs in two dEs. 
We summarize their properties in Table\,\ref{Table:nGCs}. 
\begin{table*}

\centering

\caption{Properties of nuclear clusters in late-type dwarf galaxies. 
         Column (1) gives the cluster ID; in columns (2), (3) and (4) the distance to 
	 the galaxy, D in Mpc, its absolute magnitude $M_{V,\rm Gal}$, and the foreground 
	 Galactic extinction $E_{(B-V)}$ are listed; columns (5) and (6) are the absolute magnitude 
	 $M_{V,\rm nGC}$ and $V-I$ color of the nGCs corrected for foreground extinction; in 
	 columns (7) through (13) are given the cluster mass (${\cal M}_\odot$), ellipticity 
	 $\epsilon$, half-light radius $r_{\rm h}$ (pc), logarithm of the concentration 
	 index $c=\log_{10}(r_{\rm t}/r_{\rm h})$, escape velocity $\upsilon_{\rm esc}$ (km/s) 
	 to the tidal radius, projected distance from the galaxy center $r_{\rm proj}$ 
	 in parsecs and $r_{\rm proj}$ normalized to $r_{\rm eff}=\sqrt{a\times b}$, 
	 where $a$ and $b$ are the galaxy semi major and minor axes.}

\label{Table:nGCs}

\begin{tabular}{lcccccccccccc}
\hline
\hline
\vspace*{0.1cm}ID & D	&$E_{(B-V)}$& $M_{V, \rm Gal}$&$M_{V, \rm nGC}$&$(V-I)_0$&$\cal M_{\rm nGC}$&$\epsilon$&$r_{\rm h}$&$c$&$\upsilon_{\rm esc}$&$r_{\rm proj}$&$r_{\rm proj}/r_{\rm eff}$\\
\vspace*{0.1cm}   & Mpc	& mag	& mag & mag & mag & $10^{5}{\cal M}_\odot$ &  & pc & & km/s & pc &  \\
\vspace*{0.1cm} (1) & (2) & (3) & (4) & (5) & (6) & (7) & (8) & (9) & (10) & (11) & (12) & (13) \\
\hline

E059-01-01 &  4.57  & 0.147 & $-14.60$ & $-9.89$ & 0.907 & 14.39 &  0.05  &  2.35 &  2.00 &  87.41 &  14.4 &  0.4\\
E223-09-06 &  6.49  & 0.260 & $-16.47$ & $-9.72$ & 0.921 & 12.31 &  0.21  &  3.51 &  2.00 &  66.14 & 326.4 &   5.0\\
E269-66-03 &  3.82  & 0.093 & $-13.89$ & $-9.99$ & 0.926 & 15.78 &  0.13  &  2.50 &  1.18 &  72.34 &   0.0 &   0.0\\
IC1959-04  &  6.05  & 0.011 & $-15.99$ & $-9.83$ & 0.968 & 13.62 &  0.08  &  2.92 &  2.00 &  76.28 & 228.9 &   0.2\\
KK197-02   &  3.87  & 0.154 & $-13.04$ & $-9.83$ & 0.932 & 13.62 &  0.11  &  2.95 &  1.48 &  66.56 &   0.1 &   0.0\\
N4163-01   &  2.96  & 0.020 & $-14.21$ & $-9.38$ & 0.915 &  8.99 &  0.09  &  1.45 &  1.48 &  77.17 & 152.7 &   8.3\\
U7369-10   & 11.59  & 0.019 & $-16.17$ &$-12.08$ & 0.824 &108.18 &  0.16  &  2.31 &  2.00 & 241.71 &   0.0 &   0.0\\
U8638-03   &  4.27  & 0.013 & $-13.69$ &$-10.35$ & 1.077 & 21.97 &  0.04  &  2.62 &  1.18 &  83.41 & 481.8 &  23.7\\
UA86-17    &  2.96  & 0.942 & $-16.13$ & $-9.67$ & 0.731 & 11.75 &  0.08  &  3.27 &  1.18 &  54.59 &  70.1 &   6.5\\
UA86-29    &  2.96  & 0.942 & $-16.13$ &$-11.16$ & 1.020 & 46.36 &  0.12  &  4.73 &  0.70 &  79.45 & 317.5 &  29.6\\

\hline
\hline
\end{tabular}

\end{table*}

The radial distribution of the GCs in low-mass galaxies in 
Figure\,\ref{Fig.selection} shows that nearly all of the brightest 
($M_V<-9$\,mag) clusters are located within R$_{\rm proj}\leq500$\,pc. 
This confirms the expectation based on dynamical friction above. 
The observed decrease of the average cluster luminosity with increasing 
R$_{\rm proj}$ implies that either dynamical friction is indeed strong 
in dwarfs, driving predominantly the more massive GCs inwards \cite[of 
which some might merge to form the nGC, see numerical simulation by][]{Vesperini00}. 
Another formation scenario is the ``biased formation'' of more massive 
GMCs in the nuclear regions of galaxies where the higher ambient pressure 
and density \citep{Elmegreen93, Blitz&Rosolowsky06} favors the formation 
of more massive star clusters \citep{Harris&Pudritz94}. Together with 
dynamical friction this greatly enhances the likelihood for the presence 
of high-mass nuclear globular clusters.

Probably a mixture of both processes can explain the observed 
$M_V$\,vs.\,$R_{\rm proj}$ distribution of the GCs in our sample galaxies.

\subsection{The unique case of UGCA\,86}
One of the 
most interesting cases in our sample of dwarf irregular galaxies is 
UGCA\,86 which hosts two nGCs. This galaxy has an absolute magnitude 
of $M_V=-16.13$\,mag \cite[$L=2.4\times10^8L_\odot$,][]{Georgiev09}, 
M$_{\rm HI}=1.6\times10^8$M$_\odot$, calculated following 
\cite{Roberts&Haynes94} from the total flux measured by \cite{Stil05} 
and a distance modulus of $D=2.96$\,Mpc derived by \cite{Karachentsev06}. 
UGCA\,86 is at a projected distance of $\sim81$\,kpc ($94\arcmin$) 
southeast of the bright ($M_V=-21.78$\,mag) Scd galaxy IC\,342, thus 
providing a system similar to the Milky Way and the Magellanic clouds, 
although the latter is closer to its dominant host galaxy.
\begin{figure*}
\epsfig{file=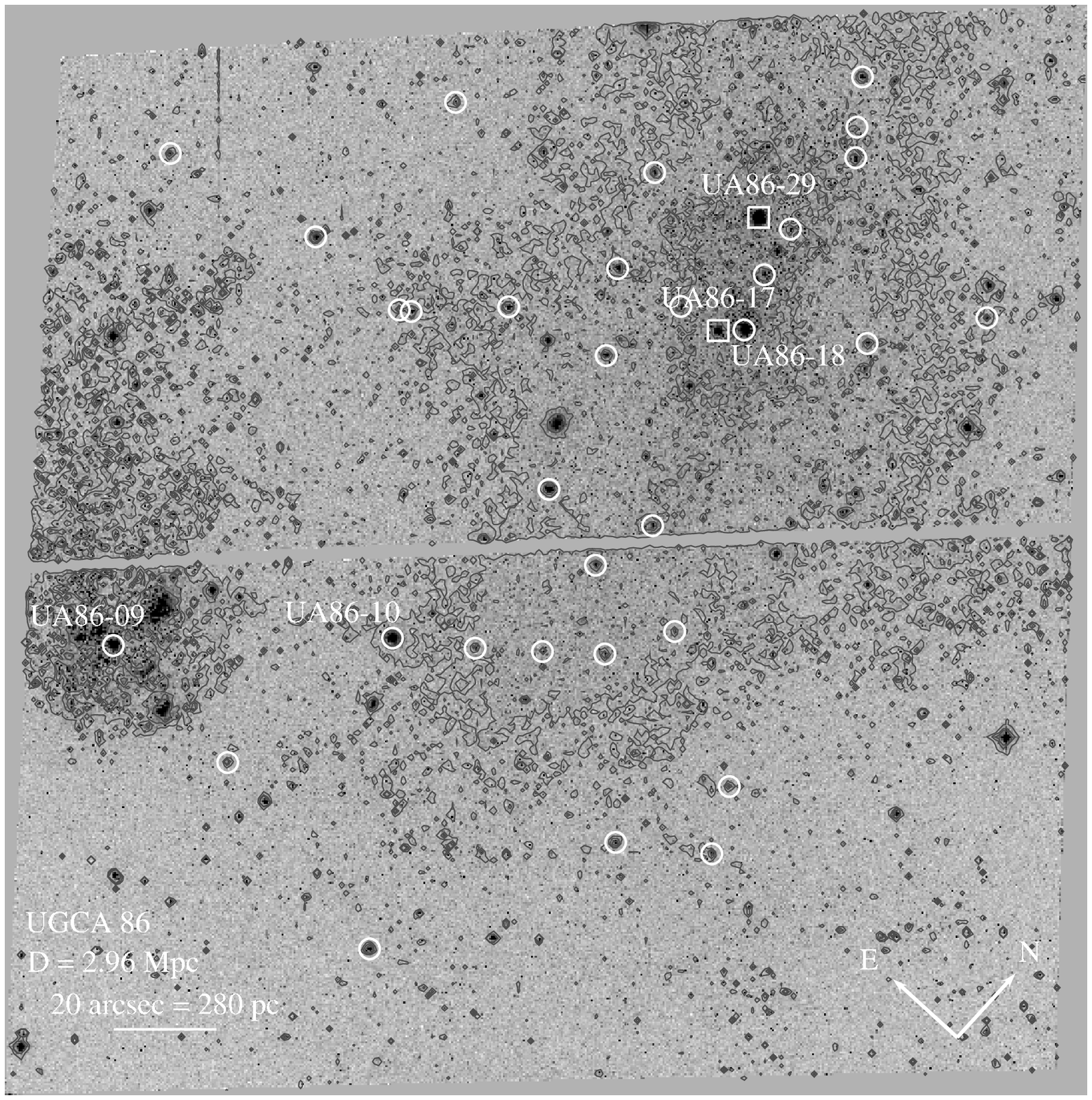, width=0.5\textwidth, bb=37 159 557 688}\epsfig{file=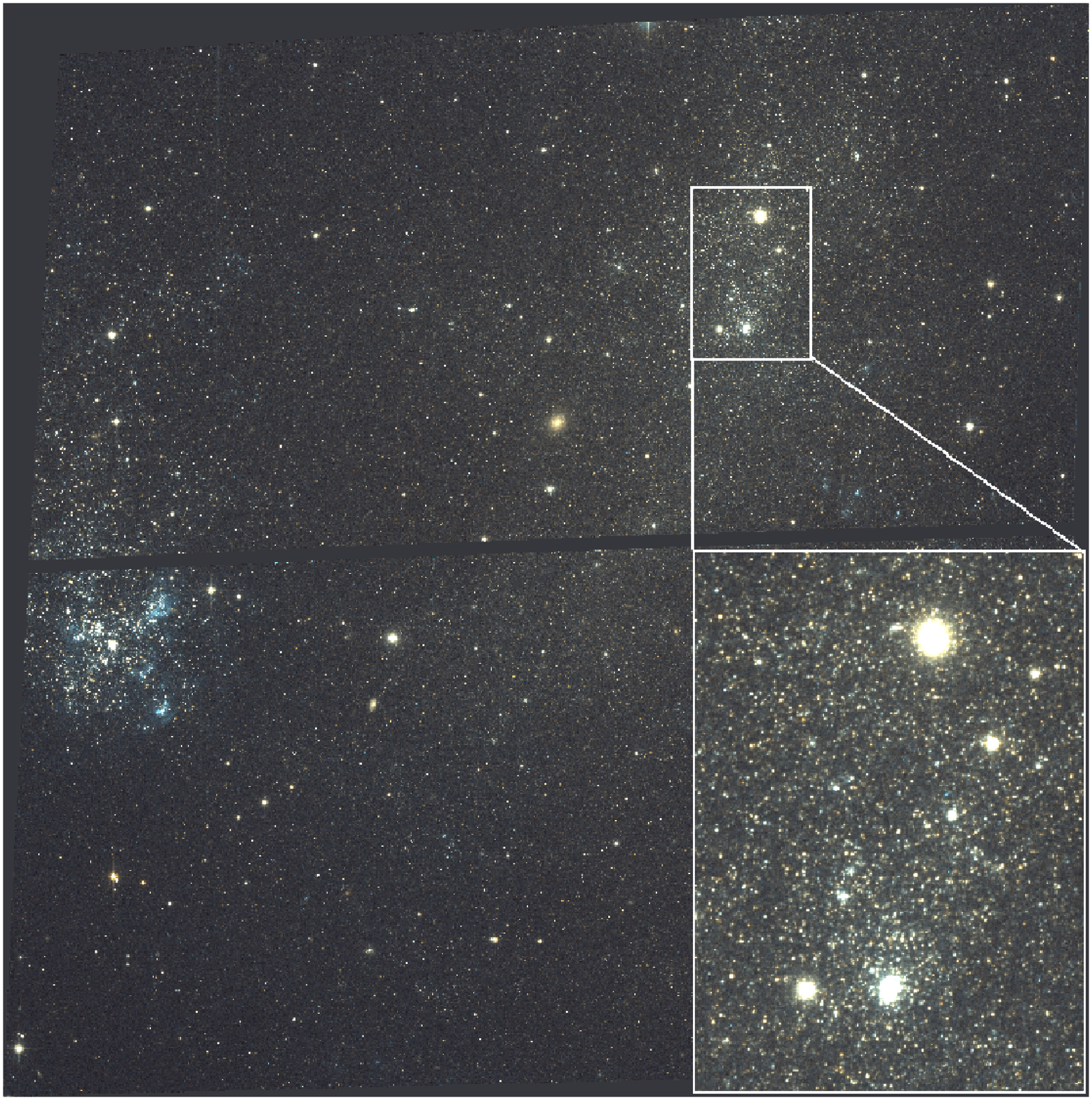, width=0.5\textwidth, bb=12 14 810 820}
\caption{{\it Left:} Gray scale F606W band HST/ACS image of UGCA\,86. 
White circles mark all GC candidates selected in \protect\cite{Georgiev09} 
while the white squares show the two clusters in the nuclear region 
of UGCA\,86. Labeled are all clusters brighter than $M_V=-9$\,mag. 
At least three more star clusters are as luminous as the nGCs, but 
have colours of $V-I<0.6$\,mag corresponding to an age of $<2$\,Gyr 
at [Fe/H$]>-2.3$\,dex. The lowest iso-intensity contour corresponds 
to $10\sigma$ above the background. {\it Right:} colour composite 
image from F606W, F814W and their average in the blue, red and green 
channels, respectively. A full-resolution image of the central 
region of UGCA\,86 is shown in the lower right corner. Both panels 
have the same scale.
\label{Fig.UA86}}
\end{figure*}
In the left panel of Figure\,\ref{Fig.UA86} we show a gray scale 
F606W-band HST/ACS image of UGCA\,86 with iso-intensity contours 
over plotted. Apart from the nGCs (white squares) this galaxy hosts 
at least three more luminous clusters, but those have $V-I$ colours consistent 
with an upper age estimate $<2$\,Gyr at [Fe/H$]>-2.3$\,dex. One of 
those bright clusters is in the nuclear region of UGCA\,86 and 
likely gravitationally interacting with the other two. Thus, we 
might be witnessing the process of merging star clusters in the 
central region of this galaxy. Given the separation ($r_i<0.4$\,kpc) 
and masses (${\cal M}\sim10^5{\cal M}_\odot$) the expected dynamical 
friction time scale is $t_{\rm df}\leq0.4$\,Gyr.

\section{Comparison between Galactic EHB-GCs and nGCs in dwarf galaxies}\label{Sect:analysis}
\subsection{Colours, luminosities and sizes}

The integrated nGC colours, luminosities and structural parameters 
from two band imaging can be used for a first order comparison of 
how similar nuclear clusters and EHB-GCs are in their metallicity 
and mass distribution.

\begin{figure}
\epsfig{file=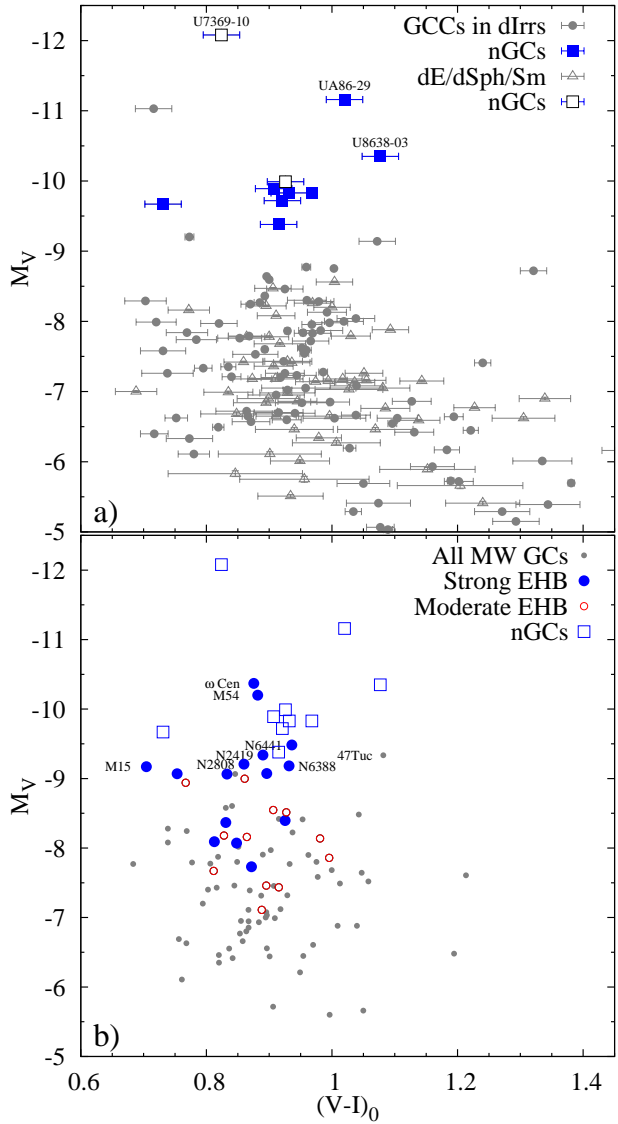, width=0.5\textwidth, bb=50 50 338 554}
\caption{colour-magnitude diagram for nuclear GCs (nGCs) in low-mass 
dwarf galaxies (squares in panels a) and b)) and Galactic GCs with 
extended horizontal branches (EHB-GCs, large circles in panel b). With 
solid and open circles in panel b) are indicated EHB-GCs with strongly 
and moderately extended HBs, respectively.
\label{Fig.CMD}}
\end{figure}
In Figure\,\ref{Fig.CMD} we compare the colour-magnitude diagrams 
(CMDs) of GCs in our sample dwarf galaxies and Galactic GCs, where 
we focus on the comparison between nGCs and EHB-GCs. In panel a) we 
show the CMD of the dwarf galaxies GCs where nGCs in dIrrs and dEs are 
shown with filled and open squares, respectively. Although the nGCs 
were selected as the brightest and the most centrally located 
clusters in our dwarf galaxies, it can be seen that they form a distinct, 
well separated group of objects at $M_V\gtrsim-9$\,mag 
($3.4\times10^5L_\odot$). Following the classification of the EHB-GCs 
by \cite{Lee07} into clusters with strongly and moderately extended HBs 
($\Delta V_{\rm HB} > 3.5$\,mag for strongly extended), in Figure\,\ref{Fig.CMD}\,b) 
we show these two groups with solid and open circles, respectively. 
It can be seen that the two EHB-GC populations show different magnitude 
distributions, with the EHB-GCs with hotter HBs being more 
luminous (and hence massive) on average. This reflects to some extent the known correlation between 
the maximum $T_{\rm eff}$ on the HB and GC mass \citep{Recio-Blanco06}. 
The open squares in Figure\,\ref{Fig.CMD}\,b) show the nGCs. They share 
the colour and magnitude distributions of the Galactic EHB-GCs. 
This suggests that nGCs have metallicity and mass distributions similar 
to the EHB-GCs, although the $V-I$ colour is rather degenerate in age and
metallicity. If we assume a lower limit   
of the cluster age of 10\,Gyr and $(V-I)=0.9$\,mag, its metallicity, 
as inferred from Simple Stellar Population (SSP) models \cite[e.g.][]{BC03} 
would be [Fe/H$]\lesssim-1.7$. However, a cluster with the same colour 
could be as metal-rich as [Fe/H$]\lesssim0.4$ and have an age of 
$\gtrsim0.7$\,Gyr, which would also significantly influence its mass 
estimate. Therefore, a solid metallicity and/or age 
indicator (from integrated-light spectra or multi-band photometry) is 
crucial and required for a robust analysis of nGC masses, ages and
metallicities. 

Numerical simulations have shown that the globular cluster half-light 
radius ($r_{\rm h}$) is a stable quantity over many cluster 
relaxation times \cite[$>10\,t_{\rm rh}$,][]{Spitzer&Thuan72, 
Aarseth&Heggie98}, and hence it represents a good proxy for the initial 
conditions in which the cluster evolved. In \cite{Georgiev09}, 
we showed that the $r_{\rm h}$ evolution of GCs in low-mass dwarf galaxies 
is mainly governed by processes internal to the cluster due to the 
weak external tidal field of the host. Therefore, a similar $r_{\rm h}$ 
distribution of nGCs and EHB-GCs would indicate an evolution 
in a similar tidal environment.
\begin{figure}
\epsfig{file=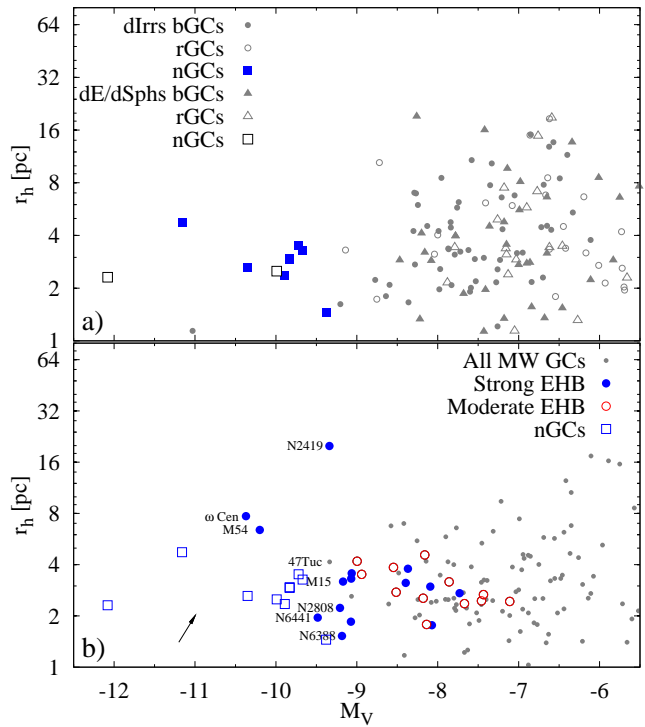, width=0.5\textwidth, bb=60 60 410 453}
\caption{Absolute magnitude vs. cluster half-light radius $r_{\rm h}$ 
for nuclear GCs (nGC) in dwarf galaxies (squares in panels a) and b)) 
and Galactic GCs with extended horizontal branches (EHB-GCs, large 
circles in panel b). The large solid and open circles in panel b) 
indicate EHB-GCs with strongly and moderately extended HBs, respectively. 
Small solid dots show all Galactic GCs. Filled and open circles and 
triangles in panel a) indicate clusters with $V-I$ colours bluer or 
redder than $(V-I)=1$\,mag. The arrow in panel b) indicates the 
expected \emph{direction} of the cluster evolution due to $r_{\rm h}$ 
expansion and mass loss from tidal stripping upon accretion.
\label{Fig.rh-M}}
\end{figure}
In Figure\,\ref{Fig.rh-M} we compare the $r_{\rm h}$\,vs.\,$M_V$ distribution 
of nGCs. The direct comparison in Figure\,\ref{Fig.rh-M}\,b) 
between nGCs and Galactic EHB-GCs shows that nGCs populate the region 
occupied mainly by the Galactic EHB-GCs with the hottest HBs, although 
they are not as extended as $\omega$\,Cen, M\,54 or NGC\,2419. Also, nGCs 
fill the gap at $M_V\sim-10$\,mag between the latter and the rest of the EHB-GCs. 

Since nGCs are in or close to the centre of the potential of their host 
galaxy, their tidal radius is relatively small. Upon accretion into 
the halo of a more massive galaxy their tidal radius $r_{\rm t}$ can 
expand, and $r_{\rm h}$ as well, while the host dwarf galaxy will 
be disrupted. This is due to the change from strong to a weaker tidal 
field in the galaxy halo after the dwarf is destroyed. This is evident 
from the identical $r_{\rm t}$ for a cluster at 0.8\,kpc in a dwarf galaxy 
and a cluster at 8\,kpc in the Galaxy \cite[Fig.\,9 in][]{Georgiev09}. 
Therefore, it is reasonable to expect that massive Galactic GCs (such as $\omega$\,Cen, 
NGC\,2419) with a large $r_{\rm h}$ at present could have evolved from 
an initially small $r_{\rm h}$ they would have had if they originated as 
nGCs in dwarf galaxies. Further, M\,54, which is still at the nucleus of 
the Sgt dSph, has a smaller $r_{\rm h}$ than $\omega$\,Cen. The masses 
of M\,54 and the stellar component of the Sgr nucleus within 100\,pc 
\cite[derived by ][]{Bellazzini08} imply a mass ratio of $\sim0.4$. 
Thus, M\,54 is still embedded in a rather deep potential well and 
its tidal radius is still confined and defined by the deeper and stronger 
potential of the Sagittarius galaxy. At the current Galactocentric 
distance of $R_{\rm GC}=19.2$\,kpc \citep{Harris96} M\,54 might undergo 
a process of $r_{\rm t}$, and therefore $r_{\rm h}$ expansion due to 
the loss of background stars (i.e. mass in its vicinity) and the resulting 
drop in cluster potential upon the disruption of the Sgr dSph. Perhaps 
NGC\,2419, the most extended Galactic cluster, went through such $r_{\rm h}$ 
expansion. Moreover, its high ellipticity \cite[$\epsilon=0.14$,][]{Bellazzini07}, 
matches the average value of the nGCs ellipticity ($\epsilon=0.11$, see 
Table\,\ref{Table:nGCs}), which further supports its external origin.

Due to tidal stripping and bulge/disk shocking a cluster can also lose 
a fraction of its mass. \cite{Ideta&Makino04} were able to reproduce 
with N-body modeling the surface brightness profile of $\omega$\,Cen if it was stripped 
from a dE host galaxy with baryon mass $1.3\times10^8{\cal M}_\odot$ 
without the need of including dark matter. As a consequence of the 
stripping, the cluster lost nearly $40\%$ of its initial mass within 
0.88\,Gyr of the stripping process. Thus, the expected direction of the cluster evolution is toward larger 
$r_{\rm h}$ and dimmer magnitude (smaller mass), as shown by the arrow 
in Figure\,\ref{Fig.rh-M}.

However, a detailed modeling is required to verify this scenario and quantify the degree of $r_{\rm h}, M_V$ evolution. It 
is also very important that the final orbit of the accreted cluster 
either is very eccentric or circular with a large galactocentric 
radius so that the cluster will spend most of the time on its orbit 
in the galactic halo.

\subsection{Internal cluster escape velocity}\label{Sect:vesc}

As discussed in the introduction, the main parameters that govern 
the HB morphology are the cluster mass, [Fe/H] and age. Here we will 
show that the escape velocity to reach the cluster tidal radius 
$\upsilon_{\rm esc}\propto\sqrt{2GM_{\rm cl}/r_{\rm h}}$, can be used 
as an additional parameter to quantify the ability of a cluster to retain 
processed material from stellar ejecta. During cluster formation, if 
retained, this chemically enriched material can be used for the formation 
of subsequent stellar populations, as recently observed in some massive GCs.

\begin{table}
 \centering
\caption{Proportionality factors taking into account the dependence of 
$\upsilon_{\rm esc}$ on the cluster density profile. In Column (1) 
the concentration index is given. Columns (2) and (3) list the correction 
factors for $\upsilon_{\rm esc}$ to reach the tidal radius and to infinity.}
\label{table:fc}

\begin{tabular}{ccc}
\hline
\hline
\vspace*{0.1cm}$\log(r_{\rm t}/r_{\rm c})$ & $f_{\rm c,\ t}$ & $f_{\rm c,\ \infty}$\\
\vspace*{0.1cm}(1) & (2) & (3)\\
\hline

  0.5 & 0.07637 & 0.09126 \\
  0.6 & 0.07852 & 0.09183 \\
  0.7 & 0.08068 & 0.09250 \\
  0.8 & 0.08285 & 0.09328 \\
  0.9 & 0.08501 & 0.09418 \\
  1.0 & 0.08717 & 0.09520 \\
  1.1 & 0.08934 & 0.09635 \\
  1.2 & 0.09154 & 0.09765 \\
  1.3 & 0.09379 & 0.09910 \\
  1.4 & 0.09610 & 0.10071 \\
  1.5 & 0.09849 & 0.10249 \\
  1.6 & 0.10098 & 0.10445 \\
  1.7 & 0.10359 & 0.10659 \\
  1.8 & 0.10634 & 0.10893 \\
  1.9 & 0.10923 & 0.11147 \\
  2.0 & 0.11229 & 0.11422 \\
  2.1 & 0.11553 & 0.11720 \\
  2.2 & 0.11897 & 0.12040 \\
  2.3 & 0.12261 & 0.12384 \\
  2.4 & 0.12647 & 0.12753 \\
  2.5 & 0.13057 & 0.13148 \\
\hline
\hline
\end{tabular}

\end{table}

In order for enriched stellar ejecta to be kept within the cluster, 
their terminal velocity $(\upsilon_\infty)$ must be no greater than 
the escape velocity to reach the cluster tidal radius $(\upsilon_{\rm esc})$. 
\cite{Leitherer92} have shown that for hot and massive OB stars 
$\upsilon_\infty$ scales with metallicity as $\upsilon_\infty \propto Z^{0.13}$, 
while for AGB stars the relation is steeper, i.e 
$\upsilon_\infty \propto Z^{0.5}\ L^{0.25}$ \cite[e.g.][]{Elitzur&Ivezic01,Marshall04}. 
A typical wind velocity of an AGB star derived from observations is $10-20$\,km/s 
\cite[e.g.][]{Vassiliadis93, Bloecker95, Habing&Olofsson03}. Rapidly rotating 
stars are known to have radial mechanical winds at 
the equator (where at the breakup limit the centrifugal force overcomes 
the gravity) with velocities from a few to a few hundred km s$^{-1}$
\citep{Decressin07b, Porter&Rivinius03}. Supernovae have winds with speed 
of thousands of km s$^{-1}$. Based on 2D hydrodynamical simulations, 
\cite{Wunsch08} have shown that thermalized ejecta from massive stars 
in compact and massive clusters are dense enough to trigger density 
condensations and feed subsequent star formation. Therefore, massive 
stars rapidly ($<50$\,Myr) deposit (through stellar winds and SNe explosions) 
a large fraction of their mass back into the immediate ISM. Due to stochastic 
effects, more massive clusters will have relatively more massive stars 
(drawn from the sparsely populated high-mass end of the IMF) than less 
massive clusters. More massive clusters also have a higher capability 
to retain stellar ejecta, which subsequently 
would lead to a higher degree of self-enrichment due to their 
high escape velocity, which is also a function of metallicity. 
Thus our primary goal is to quantify and test the viability of such 
a scenario by looking at the escape velocity to reach the cluster's tidal 
radius. We use the following expression to calculate $\upsilon_{\rm esc}$:
\begin{equation}\label{eqn:vesc}
\upsilon_{\rm esc}=f_{\rm c}\sqrt{\frac{M_{\rm cl}}{r_{\rm h}}}\ \ {\rm [km/s]},
\end{equation}
where $r_{\rm h}$ and $M_{\rm cl}$ are the cluster half-light radius 
in parsecs and cluster mass in $M_\odot$; $f_{\rm c}$ is a coefficient 
which takes into account the dependence of the escape velocity on 
the density profile of the cluster, i.e. its concentration $c=\log(r_{\rm t}/r_{\rm c})$. 
This coefficient was computed for \cite{King62} models by de-projecting 
the density profile and then calculating the potential as a function 
of radius. We calculated both, the cluster velocity needed to reach 
the cluster tidal radius and the one to reach infinity. The values 
we have used for $f_{\rm c}$ are those at the tidal radius. They differ 
insignificantly from the values at infinity, i.e. leading to 
$\Delta\upsilon_{\rm esc}<1$\,km/s. Both $f_{\rm c}$ values are 
listed in Table\,\ref{table:fc}.
Calculating $\upsilon_{\rm esc}$ involves knowledge of the 
cluster's half-light radius, mass and concentration. All these 
quantities are easily accessible from ground and space based 
observations of GCs in nearby galaxies. Given that the cluster 
mass is the quantity that varies more significantly (an order of 
magnitude) than $r_{\rm h}$ among clusters, $\upsilon_{\rm esc}$ 
mainly traces cluster mass.

The $r_{\rm h}$ measurement of the GCs in our dwarfs is described 
in detail in \cite{Georgiev08,Georgiev09}. 
To compute cluster masses we have used the luminosities we measured 
in \cite{Georgiev08,Georgiev09} and a $M/L=1.88$, which is the mean 
$M/L$ value for old LMC GCs \cite[estimated from the measurements of]
[]{McLaughlin&vdMarel05}. The basic properties and $\upsilon_{\rm esc}$ 
of all GC candidates in our sample dwarf galaxies are provided in 
Table\,\ref{Table:All_GCs_short}.
\begin{table*}
\caption{Properties of nuclear clusters of late-type dwarf galaxies. 
         In Column (1) we give the cluster ID; in (2) the distance to the galaxy D adopted 
	 from \protect\cite{Tully06,Karachentsev06,Karachentsev07}; (3) foreground Galactic 
	 extinction E(B-V) from NED; in columns (4) and (5) the cluster $V-I$ color 
	 (corrected for foreground extinction) and its absolute magnitude $M_V$ are given; in columns 
	 (6) through (12) are the logarithm of the cluster mass ${\cal M}$, ellipticity 
	 $\epsilon$, half-light radius $r_{\rm h}$, logarithm of the concentration index 
	 $c=\log(r_{\rm t}/r_{\rm c})$, escape velocity to reach the tidal radius $\upsilon_{\rm esc}$, 
	 projected distance from the galaxy center $r_{\rm proj}$ and $r_{\rm proj}$ 
	 normalized to galaxy $r_{\rm eff}=\sqrt{a\times b}$, where $a$ and $b$ are the 
	 galaxy semi major and minor axes. The table is divided in several parts to reflect the 
	 different galaxy morphology.
	 \label{Table:All_GCs_short}}

\begin{tabular}{lccccccccccc}
\hline
\hline
ID & D & E(B-V) & $(V-I)_0$ & $M_V$ & $\log_{10}\cal{M}$ & $\epsilon$ & $r_{\rm h}$ & $c$ & $\upsilon_{\rm esc}$ & $r_{\rm proj}$ & $r_{\rm proj}/r_{\rm eff}$ \\
 & Mpc & mag & mag & mag & $\cal{M}_\odot$ &  & pc &  & km/s & pc & \\
(1) & (2) & (3) & (4) & (5) & (6) & (7) & (8) & (9) & (10) & (11) & (12) \\

\hline

&  &  &  &  &  &  &  &  &  &  &  \\
\multicolumn{12}{c}{dIrrs}\\
&  &  &  &  &  &  &  &  &  &  &  \\

D565-06-01  &  9.08 & 0.039 & 1.583&$  -6.10$&   4.64&   0.36&    1.19&   2.00&  21.44&  524.2&   13.1 \\
D634-03-01  &  9.55 & 0.038 & 1.039&$  -7.08$&   5.03&   0.01&    5.93&   0.70&  10.84&  298.4&    3.4 \\
DDO52-01    & 10.28 & 0.037 & 1.004&$  -6.62$&   4.85&   0.03&    3.34&   1.18&  13.26&  840.5&   14.3 \\
\nodata & \nodata & \nodata & \nodata & \nodata & \nodata & \nodata & \nodata & \nodata & \nodata & \nodata & \nodata \\

&  &  &  &  &  &  &  &  &  &  &  \\
\multicolumn{12}{c}{Sms}\\
&  &  &  &  &  &  &  &  &  &  &  \\

E137-18-01  &  6.40 & 0.243 & 1.030&$  -7.79$&    5.32&   0.03&    3.42&   2.00&  27.55&  654.3&   14.8 \\
E274-01-07  &  3.09 & 0.257 & 0.945&$  -6.86$&    4.95&   0.00&     2.8&   1.18&  16.17&  616.4&   25.8 \\
N247-02     &  3.65 & 0.018 & 1.138&$  -6.59$&    4.84&   0.09&   18.79&   1.48&   5.93& 1395.9&   27.2 \\
N4605-10    &  5.47 & 0.014 & 0.969&$  -8.26$&    5.51&   0.01&   19.16&   2.00&  14.45& 2938.6&   64.0 \\
\nodata & \nodata & \nodata & \nodata & \nodata & \nodata & \nodata & \nodata & \nodata & \nodata & \nodata & \nodata \\

&  &  &  &  &  &  &  &  &  &  &  \\
\multicolumn{12}{c}{dSphs}\\
&  &  &  &  &  &  &  &  &  &  &  \\

IKN-01      &  3.75 & 0.061 & 0.911&$  -6.65$&    4.86&   0.01&    6.62&   0.70&   8.42&  399.2&  \nodata \\
KKS55-01    &  3.94 & 0.146 & 0.907&$  -7.36$&    5.15&   0.01&    4.51&   1.18&  16.04& 1366.6&   85.3 \\
\nodata & \nodata & \nodata & \nodata & \nodata & \nodata & \nodata & \nodata & \nodata & \nodata & \nodata & \nodata \\

&  &  &  &  &  &  &  &  &  &  &  \\
\multicolumn{12}{c}{dEs}\\
&  &  &  &  &  &  &  &  &  &  &  \\

E269-66-01  &  3.82 & 0.093 & 0.911&$  -8.08$&    5.43&   0.02&    2.87&   1.48&  30.14&  492.9&   24.1 \\
U7369-01    & 11.59 & 0.019 & 0.899&$  -6.90$&    4.96&   0.04&    2.77&   1.48&  17.82& 3430.3&   54.7 \\
\nodata & \nodata & \nodata & \nodata & \nodata & \nodata & \nodata & \nodata & \nodata & \nodata & \nodata & \nodata \\
\hline
\hline

\end{tabular}
\end{table*}

For the majority (85) of the Galactic GCs we have used the most 
recent measurements by \cite{McLaughlin&vdMarel05} of their $r_{\rm h}$, 
concentration and $M/L$ ratio, which is needed to estimate $M_{\rm cl}$. 
For the remaining GCs we adopted the values in the \cite{Harris96} 
catalog and $M/L=2.0$. Table\,\ref{Table:MW_short} 
contains the basic properties 
\begin{table*}

\caption{Properties of Galactic globular clusters. \newline
         In columns list the cluster Name in (1); foreground Galactic extinction E(B-V) from \protect\cite{Harris96} 
	 (2); corrected for foreground extinction cluster $V-I$ color in (3) and its 
	 absolute magnitude $M_V$ in (4); logarithm of the cluster mass $\cal{M}$ in 
	 (5), half-light radius $r_{\rm h}$ in (6), logarithm of the concentration 
	 index $c=\log(r_{\rm t}/r_{\rm c})$ in (7), escape velocity $\upsilon_{\rm esc}$ 
	 in (8), metallicity [Fe/H] in (9), HB ratio HBR from \protect\cite{Harris96} 
	 in (10), HB morphology in (11) sEHB and mEHB, B and R for strong and moderately 
	 extended \protect\cite[from][]{Lee07}, blue and red HBs, if HBR$>0$ or HBR$<0$, 
	 respectively. In column (12) is listed the classification of the cluster (Class) 
	 according to which Galactic GC sub-population it belongs: buldge/disk (BD), 
	 old and young halo (OH) and (YH), Saggitarius (SG) or with unknown classification 
	 (UN). The cluster Class was adopted from \protect\cite{Mackey&vdBergh05}, 
	 which follows the original \protect\cite{Zinn93} classification. 
	 \label{Table:MW_short}}

\begin{tabular}{lccccccccccc}
\hline
\hline
Name & E(B-V) & $(V-I)_0$ & $M_V$ & $\log_{10}\cal{M}$ & $r_{\rm h}$ & $c$ & $\upsilon_{\rm esc}$ & [Fe/H] & HBR & HB morphology & Class \\
 & mag & mag & mag & $\cal{M}_\odot$ & pc & & km/s & dex &  &  & \\
(1) & (2) & (3) & (4) & (5) & (6) & (7) & (8) & (9) & (10) & (11) & (12)\\
\hline
NGC5139/$\omega$Cen  &  0.12&	0.88	&	$ -10.37$&   6.37&   7.71&   0.98&  53.96	&	$  -1.62$&	\nodata& sEHB&   UN \\
NGC6715/M54          &  0.15&	0.88	&	$ -10.20$&   6.29&   6.40&   1.09&  71.01	&	$  -1.58$	&	$   0.75$& sEHB&   SG \\
NGC6441              &  0.47&	0.94	&	$  -9.48$&   6.16&   1.95&   1.11&  94.87	&	$  -0.53$&  \nodata& sEHB&   BD \\
NGC2419              &  0.11&	0.89	&	$  -9.34$&   5.95&  19.91&   1.01&  21.09	&	$  -2.12$	&	$   0.86$& sEHB&   OH \\
NGC104/47Tuc         &  0.04&	1.08	&	$  -9.34$&   6.05&   4.15&   1.09&  65.92	&	$  -0.76$	&	$  -0.99$& R   &   BD \\
NGC2808              &  0.22&	0.86	&	$  -9.21$&   5.93&   2.23&   1.07&  63.96	&	$  -1.15$	&	$  -0.49$& sEHB&   OH \\
NGC6388              &  0.37&	0.93	&	$  -9.18$&   6.02&   1.52&   1.11&  90.73	&	$  -0.60$&  \nodata& sEHB&   BD \\
\nodata & \nodata & \nodata & \nodata & \nodata & \nodata & \nodata & \nodata & \nodata & \nodata & \nodata & \nodata \\

&  &  &  &  &  &  &  &  &  &  &  \\
\multicolumn{12}{c}{Part 2}\\
&  &  &  &  &  &  &  &  &  &  &  \\

AM4                  &  0.04&  \nodata	&	$  -1.60$&   2.87&   3.65&   0.50&   1.09	&	$  -2.00$&  \nodata& \nodata  &   UN \\
Djorg1               &  1.44&  \nodata	&	$  -6.26$&   4.73&   4.40&   1.50&  10.97	&	$  -2.00$&  \nodata& \nodata  &   UN \\
Djorg2/ESO456-SC38   &  0.89&  \nodata	&	$  -6.98$&   5.02&   1.62&   1.50&  25.19	&	$  -0.50$	&	$  -1.00$& R   &   BD \\
E3                   &   0.3&  \nodata	&	$  -2.77$&   3.34&   2.58&   0.75&   2.35	&	$  -0.80$&  \nodata& \nodata  &   UN \\
Eridanus             &  0.02&  \nodata	&	$  -5.14$&   4.29&  10.50&   1.10&   3.85	&	$  -1.46$	&	$  -1.00$& R   &   YH \\
HP1/BH229            &  0.74&  \nodata	&	$  -6.44$&   4.81&   6.20&   2.50&  13.31	&	$  -1.55$&  \nodata& \nodata  &   OH \\
Liller1              &  3.06&  \nodata	&	$  -7.63$&   5.28&   1.26&   2.30&  47.96	&	$   0.22$	&	$  -1.00$& R   &   BD \\
NGC4372              &  0.39&	0.93	&	$  -7.77$&   5.34&   6.58&   1.30&  17.12	&	$  -2.09$	&	$   1.00$& B   &   OH \\
NGC4833              &  0.32&	0.86	&	$  -8.16$&   5.49&   4.56&   1.25&  24.02	&	$  -1.80$	&	$   0.93$& mEHB&   OH \\
\nodata & \nodata & \nodata & \nodata & \nodata & \nodata & \nodata & \nodata & \nodata & \nodata & \nodata & \nodata \\
\hline
\hline	     
\end{tabular}
\end{table*}

of Galactic globular clusters compiled from \cite{Harris96} and 
\cite{McLaughlin&vdMarel05} used to calculate $\upsilon_{\rm esc}$. 
The first part of the table contains the $M_{V}$, $\cal{M}$, $r_{\rm h}$, $c$ 
and $\upsilon_{\rm esc}$ for 85 GCs calculated or adopted from 
\cite{McLaughlin&vdMarel05}. The rest of the GCs (second part of 
the table) are from the entries in the Harris catalog supplemented 
with $r_{\rm h}$ from \cite{Mackey&vdBergh05}. We have updated the 
metallicity value of NGC\,6440 from \cite{Origlia08}.

\subsection{$\upsilon_{\rm esc}-$metallicity relation}

In the light of the self-enrichment scenario, clusters have to retain 
enriched stellar ejecta from their first stellar population 
in order to show the observed abundance anomalies and/or multiple 
horizontal branches. Therefore, more massive (higher $\upsilon_{\rm esc}$) 
cluster are likely to retain a larger amount of processed stellar 
material which can also lead to an increase of their (initial) metallicity. 
Therefore, we expect an $\upsilon_{\rm esc}$ (mass)-metallicity relation. 
We should note, however, that a recent model by \cite{Bailin&Harris09} 
of the mass--metallicity relation (MMR) among the blue GCs in rich 
GC systems of massive galaxies shows that the MMR can not 
arise from self-enrichment, but rather from pre-enrichment of the 
GMC. However, their model did not include feedback from stars with 
slow stellar winds and enhanced mass loss such as AGB stars and fast 
rotators. In the following, we will show that when the $\upsilon_{\rm esc}$ 
is considered, such a modified MMR can actually be seen for Galactic EHB-GCs.
\begin{figure}
\epsfig{file=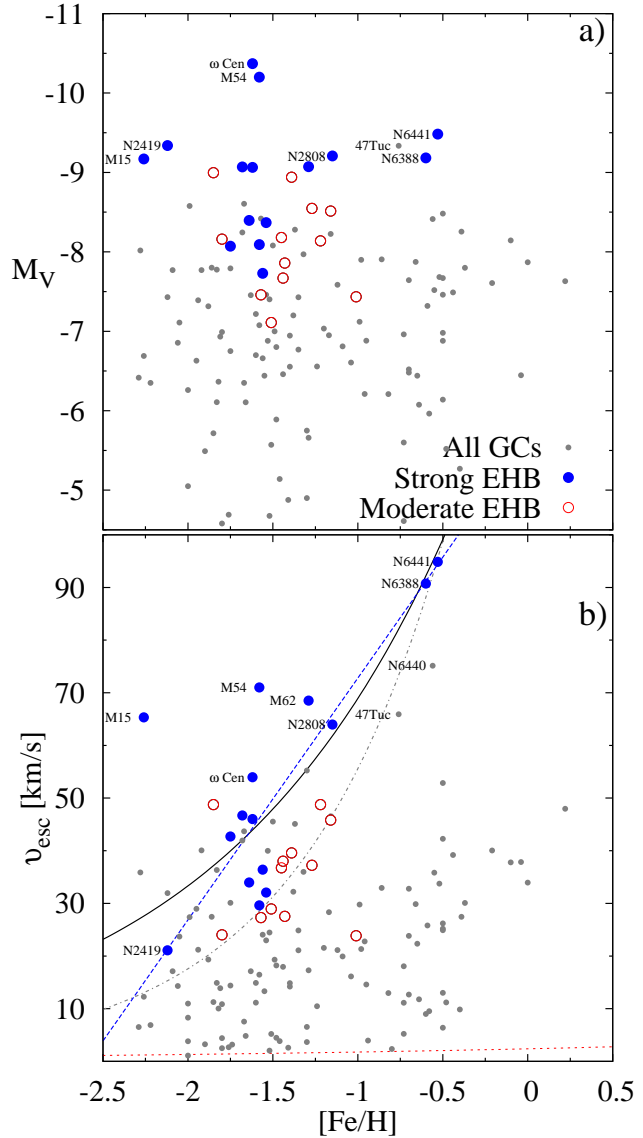, width=0.5\textwidth, bb=50 50 338 554}
\caption{Metallicity vs. cluster luminosity (panel a) and internal cluster 
escape velocity $\upsilon_{\rm esc}$ (panel b) for Galactic GCs. In both 
panels, large solid and open circles indicate Galactic GCs with strongly 
or moderately extended horizontal branches (EHB-GCs), respectively. 
The dashed line in panel b) shows a linear least-square fit to the EHB-GCs 
excluding M\,15. With a dash-dotted curve is shown the relation 
between terminal velocity and metallicity for a dust-driven AGB wind 
model ($\upsilon_\infty \propto Z^{0.5}\ L^{0.25}$). The solid curve 
indicates the least square fit to the EHB-GCs, which resulted in a 
flatter relation $\upsilon_\infty\propto Z^{0.32}$. The dotted curve 
shows the $\upsilon_{\infty}\propto Z^{0.13}$ relation from \protect\cite{Leitherer92} 
for an AGB star.
\label{Fig.Mv-Z-vesc}}
\end{figure}
In Figure\,\ref{Fig.Mv-Z-vesc}\,a) we show the metallicity versus 
cluster luminosity/mass for Galactic GCs. All EHB-GCs are among the 
brightest, i.e. most massive clusters, and are typically metal-poor.
The exceptions are NGC\,6441 and NGC\,6388, which have higher metallicities. 
However, 47\,Tuc, which is (at least) as massive as most of the EHB-GCs 
and as metal-rich as NGC\,6441 and NGC\,6388, doesn't show an extended HB.

In Figure\,\ref{Fig.Mv-Z-vesc}\,b), we show 
[Fe/H] vs.\ $\upsilon_{\rm esc}$ for Galactic GCs. With small dots, 
large solid and open circles are shown all Galactic GCs and EHB-GCs 
with strong or moderate extension of the HB, respectively. It can 
be seen that there is a correlation between these two quantities 
for EHB-GCs. The dashed line shows the least-square fit to the 
EHB-GCs excluding M\,15, which clearly stands out from the 
rest of the EHB-GCs. M\,15 is a core collapsed cluster \cite[e.g.][and 
refs therein]{Guhatakhurta96, Noyola&Gebhardt06}, which could have 
experienced a strong $r_{\rm h}$ evolution. Such expansion of its 
$r_h$ after core collapse occurs due to binary heating in the center 
\citep{Baumgardt02}, resulting in a low {\it present-day} $\upsilon_{\rm esc}$. 
Thus, the {\it initial} $\upsilon_{\rm esc}$ of M\,15 was even higher. In 
addition, a King profile does not represent well the surface brightness 
profile of a post-core collapse cluster \citep{Trager95}, thus the 
$f_c$ we used to calculate $\upsilon_{\rm esc}$ for $c=2.5$ from 
the \citeauthor{Harris96} catalog might be underestimated. 
\cite{Pasquali04} performed a simultaneous fit to the surface brightness 
and velocity dispersion profile of M\,15 using a multi-mass King-Michie 
model. Using the values they obtain for $r_c$ and $r_t$, we estimated 
that M\,15 has slightly larger concentration, $c=2.65$. This implies 
an increase in $\upsilon_{\rm esc}$ of about 10\,km/s from its value 
calculated with $c=2.5$ (cf. Table\,\ref{table:fc}). Leaving the details 
about M\,15 aside, we note that clusters to the left of the 
model lines are expected to be in the self-enrichment regime, which is 
consistent with the observed peculiarities of the EHB-GCs.

The observation that the higher the cluster $\upsilon_{\rm esc}$, the 
more metal-rich it is, may reflect the metallicity dependence of the 
terminal velocities of the stellar winds. The 
$\upsilon_{\rm esc}$ of a metal-rich cluster must be higher in 
order to retain such fast winds.  We suggest that 
this might be the reason why the majority of the metal-rich Galactic 
GCs do not show EHBs while being as massive as the metal-poor 
EHB-GCs. 
With a dotted curve in Figure\,\ref{Fig.Mv-Z-vesc}\,b we show the 
relation between the stellar wind terminal velocity and metallicity 
for an AGB star with $T_{\rm eff}=3000$\,K, ${\cal M}=5{\cal M_\odot}$ 
and $L=10^4L_\odot$ using equation 2 in \cite{Leitherer92}. It is  
clear that the $\upsilon_\infty\propto Z^{0.13}$ proportionality is 
not as steep as the observed relation for $\upsilon_{\rm esc}-$[Fe/H] 
for EHB-GCs. 
However, the Leitherer model is designed for studying 
radiative winds of hot ($T_{\rm eff}>15000$\,K) and massive 
(${\cal M}>15{\cal M}_\odot$) stars, which have stellar winds 
$>1000$\,km/s and different mass loss rate. Thus this wind model 
is inappropriate in reproducing the wind properties of AGB stars. 
With a dash-dotted line is shown the dust-driven AGB wind model 
of the form $\upsilon_\infty \propto Z^{0.5}\ L^{0.25}$ 
\citep{Marshall04}. The vertical normalization, which is a complex 
parameter reflecting the relation between the gas-to-dust ratio and $Z$ 
\cite[see][]{Elitzur&Ivezic01}, in this case is adopted as such to approximate 
the EHB-GCs distribution. With a solid curve in Fig.\,\ref{Fig.Mv-Z-vesc} 
we show a least square fit to the EHB-GCs resulting in a flatter relation 
$\upsilon_\infty\propto Z^{0.32}$.

It is interesting to note that the Galactic GC NGC\,6440 has very 
high present day $\upsilon_{\rm esc}$ and a high metallicity \cite[${\rm Fe/H}=-0.56$,][]{Origlia08}, 
but is classified as cluster with a red HB. NGC\,6440 is located 
in the inner Galactic bulge at $(l,b)=(7\fdg7,3\fdg8)$, i.e. 
$R_{\rm GC}=1.3$\,pc and $R_\odot=8.4$\,pc \citep{Harris96} and 
has a very high foreground extinction $A_V\simeq3.5$\,mag, i.e. 
reddening of $E(B-V)=1.15$ \citep{Valenti07}. Its HST/WFPC2 CMD 
\cite[Fig.\,4 in][]{Piotto02} shows a well populated red clump, 
however, the expected region of the extension of the HB (subluminous 
blue stars) falls in the magnitude range where the photometric 
errors are larger than 0.1\,mag, likely affected by the high 
reddening. Thus, the HB classification of NGC\,6440 might not 
be very well established.

\begin{figure}
\epsfig{file=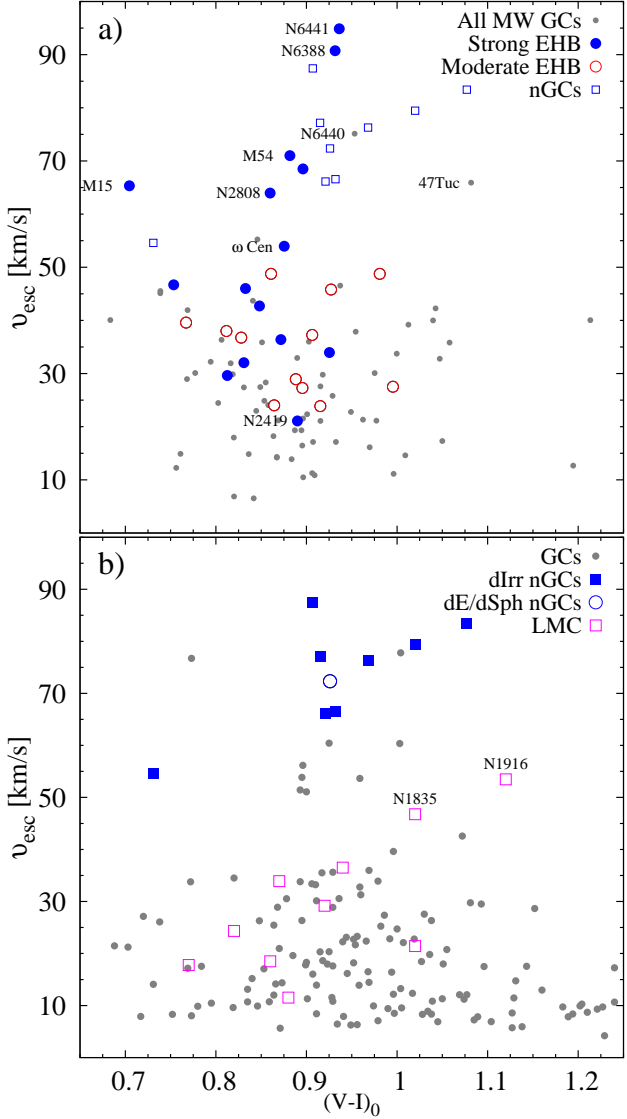, width=0.5\textwidth, bb=50 50 338 554}
\caption{Internal cluster escape velocities ($\upsilon_{\rm esc}$) versus 
cluster $V-I$ colour for Galactic GCs (panel a) and GCs in dwarf galaxies 
(panel b). With large solid and open circles in panel a) are indicated 
Galactic GCs with strongly or moderately extended horizontal branches 
(EHB-GCs), respectively. With solid squares in panel b) are shown the 
nGCs while the open squares show old LMC GCs. The nucleus of the dE 
UGC\,7369 is off in panel b) at $\upsilon = 242$\,km/s and $(V-I)_0=0.824$\,mag.
\label{Fig.VI-vesc}}
\end{figure}

Unfortunately, we do not have metallicity measurements for the 
nuclear clusters in our sample dwarfs to perform a direct comparison 
in the same parameter space between nGCs and EHB-GCs. This will 
require a spectroscopic follow-up. The closest parameter pair that 
corresponds to [Fe/H] vs. $\upsilon_{\rm esc}$ is $V-I$ vs. 
$\upsilon_{\rm esc}$, which we show in Figure\,\ref{Fig.VI-vesc}.
As can be seen, nGCs coincide with the Galactic EHB-GCs in the 
$(V-I)-\upsilon_{\rm esc}$ parameter space. Although the $V-I$ 
colour suffers from the age-metallicity degeneracy and its conversion to 
[Fe/H] is nonlinear depends on the choice of the adopted 
SSP model, the [Fe/H$]-\upsilon_{\rm esc}$ trend seen in Fig\,\ref{Fig.Mv-Z-vesc}\,b) 
can also be noticed for the Galactic EHB-GCs in Fig.\,\ref{Fig.VI-vesc}\,a). 
This is because Galactic GCs have a relatively small age spread 
(~1--2\,Gyr), therefore the $V-I$ colour will mainly trace the cluster 
metallicity. Previous studies, based on Galactic and/or GC data in 
other galaxies, derived linear relation between the $V-I$ colour 
and metallicity \cite[e.g.][]{Kundu&Whitmore98, Kissler-Patig98, 
Harris00, Sohn06}. Our nGCs, however, have an unknown age-metallicity 
composition which is likely complex \citep{Walcher06, Puzia&Sharina08}. 
Thus, combined with the low number statistics, a linear transformation 
of their $V-I$ colour to metallicity would lead to an unreliable result, 
and therefore we do not present this exercise here.

Note that $\upsilon_{\rm esc}$ in Equation\,\ref{eqn:vesc} is the 
present-day value, and hence a lower estimate of the initial $\upsilon_{\rm esc}$. 
To compare $\upsilon_{\rm esc}$ of a cluster directly with 
the terminal speed of the stellar ejecta as a function of metallicity, 
one has to take 
the evolution of the GC mass and $r_{\rm h}$ with time into account. 
Due to stellar evolution a GC loses about 40\% of its initial mass and the 
increase of $r_{\rm h}$ due to mass loss from stellar evolution is 
also about 40\% if $r_{\rm h}<\!<r_{\rm t}$ \cite[e.g.,][]{Hills80, 
Baumgardt&Makino03}. These two effects would increase the $\upsilon_{\rm esc}$ 
by a factor of 1.5 since the cluster was formed. 
However, the exact mass and $r_{\rm h}$   
evolution of a given cluster is determined by its orbital parameters 
in the galaxy and the profile of its potential. The fact that the present-day 
$\upsilon_{\rm esc}$ of EHB-GCs and nGCs are very comparable (cf. 
Fig.\,\ref{Fig.VI-vesc}) indicates that they may well have 
experienced mass and size evolution in initially similar environments. 
To provide stronger constraints on this scenario, we suggest that accurate 
ages of the nGCs shall be determined by means of follow-up spectroscopy or 
near-IR imaging.  

\section{Conclusions}\label{Sect:Conclusions}

In order to address the hypothesis that massive peculiar Galactic 
GCs, mainly those with hot horizontal branches (EHB-GCs), originated 
from nuclear clusters of accreted and now disrupted dwarf galaxies or 
were the former cores of massive Galaxy building blocks, we have performed 
a comparison of their properties with that of nuclear clusters in 
low-mass dwarf galaxies, mainly late-type irregulars. We introduce 
the escape velocity to reach the tidal radius of a cluster 
$(\upsilon_{\rm esc})$ as an additional tool to measure the ability of a 
cluster to retain enriched stellar ejecta which are required in the 
self-enrichment scenario to explain the complexity of the stellar populations 
recently observed in Galactic GCs.

The sample of nuclear clusters was presented in \cite{Georgiev08,Georgiev09} 
and is based on two band F606W and F814W 
archival HST/ACS imaging. Because the $V-I$ colour of those clusters 
is consistent with an age of $>4$\,Gyr, i.e. likely typical old GCs, 
we have termed them nuclear GCs (nGCs). Based on dynamical friction-related arguments, we selected 
eight nGCs in seven dIrr and two nGCs in two dEs with $M_V<-9$\,mag 
within projected galactocentric distances of R$_{\rm proj}<500$\,pc (cf. Table\,\ref{Table:nGCs}). 
Those clusters could have formed either at their present location 
or spiraled in due to dynamical friction within 4\,Gyr. The radial distributions 
of all GCs shown in Figure\,\ref{Fig.selection} indicates a trend 
of increasing average cluster magnitude with decreasing R$_{\rm proj}$. We 
discuss that such a relation could arise due to two mechanisms: dynamical 
friction, shown to be strong in low-mass galaxies \citep{Vesperini00}, 
or biased massive cluster formation toward lower R$_{\rm proj}$ 
reflecting the dependence of the size of giant molecular clouds 
(GMCs) on the ambient pressure, which increases with decreasing 
R$_{\rm proj}$.

We observe that the nGCs seem to form a distinct population of clusters 
being brighter than $M_V<-9$ mag and sharing a similar colour and luminosity 
distribution as the EHB-GCs. The similar $V-I$ colour between nGCs and 
EHB-GCs also indicates that they have 
similarly low metallicities, although this needs to be confirmed with 
spectroscopy and/or near-IR imaging due to the strong age-metallicity degeneracy 
of the $V-I$ colour. 

The nuclear GCs in our sample dwarfs are distributed in the $r_{\rm h}$ 
vs. $M_V$ plane in a region which connects the majority of the EHB-GCs 
and the most extended ones such as $\omega$\,Cen, NGC\,2419 and M\,54, 
the nucleus of the Sagittarius dwarf spheroidal. If such Galactic clusters had 
their origin as nGCs, it is expected that their $r_{\rm h}$ will 
suffer expansion upon accretion and disruption of the dwarf due to 
mass loss resulting in a drop of the cluster potential, i.e. the 
change from strong (in the dwarf nuclear region) to a weaker 
(in the galaxy halo) gravitational potential. This could be the reason for the 
large present-day $r_{\rm h}$ of those Galactic EHB-GCs and the 
relatively smaller ones (on average) of the nGCs in our dwarf 
galaxies. A detailed modeling of this mechanism is required in 
order to quantify the cluster $r_{\rm h}$ expansion.

Another indicator that EHB-GCs could have originated in the cores 
of dwarf galaxies is that of their chemical peculiarities, 
which could arise if they were capable to retain processed 
stellar ejecta of fast evolving massive stars. We showed that the escape 
velocity to reach the cluster tidal radius for EHB-GCs scales with cluster 
[Fe/H] in a similar manner (though much steeper) as the relation between 
the terminal velocity of the stellar wind and metallicity.
Thus, a metal-poor cluster will have better capability to retain 
stellar ejecta than a metal-rich cluster of the same mass. 

Due to the lack of metallicity measurements of the nGCs in our sample 
we compare them with the EHB-GCs in the $V-I$ vs. $\upsilon_{\rm esc}$ 
parameter space (Fig.\,\ref{Fig.VI-vesc}). They are found to occupy 
region with similar in colour and $\upsilon_{\rm esc}$ as EHB-GCs which 
indicates that nGCs could have retained ejecta from massive and intermediate-mass 
stars as did EHB-GCs. Therefore, nGCs can have as complex stellar populations 
as the EHB-GCs. However, more robust age and metallicity estimates 
for nGCs are required to confirm this.

\section*{Acknowledgments}

The authors would like to thank the referee, Prof. R.\,Gratton, for his 
constructive report which improved the discussion in the paper. IG would 
like to thank for the support for this work the German Research 
Foundation (\emph{Deut\-sche For\-schungs\-ge\-mein\-schaft, DFG\/}) 
through project number BO-779/32-1. THP acknowledges support in form 
of the Plaskett Research Fellowship at the Herzberg Institute of 
Astrophysics. The authors would like to thank Pavel Kroupa and Thibaut 
Decressin for valuable discussions and comments.

\label{lastpage}
\bibliographystyle{mn2e}
\bibliography{references}

\end{document}